\documentclass{article}

\usepackage{PRIMEarxiv}

\usepackage[utf8]{inputenc} 
\usepackage[T1]{fontenc}    
\usepackage{hyperref}       
\usepackage{url}            
\usepackage{booktabs}       
\usepackage{amsfonts}       
\usepackage{nicefrac}       
\usepackage{microtype}      
\usepackage{lipsum}
\usepackage{fancyhdr}       
\usepackage{graphicx}       
\graphicspath{{media/}}     

\usepackage{amsmath} 
\usepackage{multirow} 
\usepackage{subcaption} 

\title{MInD: Improving Multimodal Sentiment Analysis via \\ Multimodal Information Disentanglement}

\author{
    Weichen Dai$^*$\\ 
    USTC \And
    Xingyu Li\footnote{Equal contribution.} \\
    Lin Gang Lab \And
    Zeyu Wang \\
    USTC \And
    Pengbo Hu \\
    USTC \AND
    Ji Qi \\
    China Mobile Limited \And
    Jianlin Peng \\
    China Mobile Limited \And
    Yi Zhou \\
    USTC
}

\begin{document}
\maketitle

\begin{abstract}
Learning effective joint representations has been a central task in multi-modal sentiment analysis. 
Previous works addressing this task focus on exploring sophisticated fusion techniques to enhance performance. 
However, the inherent heterogeneity of distinct modalities remains a core problem that brings challenges in fusing and coordinating the multi-modal signals at both the representational level and the informational level, impeding the full exploitation of multi-modal information. 
To address this problem, we propose the Multi-modal Information Disentanglement (MInD) method, which decomposes the multi-modal inputs into modality-invariant and modality-specific components through a shared encoder and multiple private encoders. 
Furthermore, by explicitly training generated noise in an adversarial manner, MInD is able to isolate uninformativeness, thus improves the learned representations. 
Therefore, the proposed disentangled decomposition allows for a fusion process that is simpler than alternative methods and results in improved performance. 
Experimental evaluations conducted on representative benchmark datasets demonstrate MInD's effectiveness in both multi-modal emotion recognition and multi-modal humor detection tasks. 
Code will be released upon acceptance of the paper.
\end{abstract}

\section{INTRODUCTION}

Recently, there has been a growing interest in Multi-modal Sentiment Analysis (MSA).
The comprehension of sentiment is often enhanced by cross-modal input, such as visual, audio, and textual information.
Consequently, researchers have focused on developing effective joint representations that integrate all relevant information from the collected data~\cite{poria2020beneath}, while most of the models rely on designing sophisticated fusion techniques~\cite{zadeh2017tensor} for the exploration of the intra-modal and inter-modal dynamics.
Although multi-modal learning has been theoretically shown to outperform uni-modal learning~\cite{huang2021makes}, in practice, the modality gap resulting from the inherent heterogeneity of distinct modalities hampers the full exploitation of the inter-modal information for effective multi-modal representations. 
This phenomenon persists across a broad range of multi-modal models, covering texts, natural images, videos, medical images, and amino-acid sequences~\cite{liang2022mind}.
Therefore, prior approaches that address the representations of each modality through a comprehensive learning framework may lead to insufficiently refined and potentially redundant multi-modal representations.

Recent studies have initiated an exploration into the learning of distinct multi-modal representations.
Pham et al.~\cite{pham2019found} translates a source modality to a target modality for joint representations using cyclic reconstruction.
Mai et al.~\cite{mai2020modality} also provides a adversarial encoder-decoder classifier framework to learn a modality-invariant embedding space through translating the distributions.
But these methods do not explicitly learn the modality-specific representations which reveal the unique characteristic of emotions from different perspectives.
By adopting the shared-private learning frameworks~\cite{bousmalis2016domain}, Hazarika et al.~\cite{hazarika2020misa} and Yang et al.~\cite{yang2022disentangled} attempt to incorporate a diverse set of information by learning different factorized subspace for each modality in order to obtain better representations for fusion. 
However, their approaches either utilizes simple constraints that fail to guarantee a perfect factorization, or relies on a complex fusion module which may indicate that the extracted information may be unrefined.
Moreover, they both neglect the control in the information flow, which could result in the loss of practical information.

Motivated by the above observations, we propose the Multi-modal Information Disentanglement (MInD) approach to deal with the insufficient exploitation of information from heterogeneous modalities.
The main strategy is to decompose features of each modality with information optimization.
Specifically, the first component is the modality-invariant component, which can effectively capture the underlying commonalities and explore the shared information across modalities.
Secondly, we train the modality-specific component to capture the distinctive information and characteristic features.
Furthermore, as unknown noise of each modality may be categorized as complementary components, we explicitly train the generated noise in an adversarial manner to enhance the refinement of the learned information and mitigate the impact of uninformativeness on the quality of the representations. 
The combination of the modality-invariant components and the modality-specific components thus enables a more straightforward fusion process compared to alternative methods, resulting in enhanced performance.

The contributions of this paper can be summarized as:
\begin{itemize}
    \item We propose MInD, a disentanglement-based multi-modal sentiment analysis method driven by information optimization.
    MInD overcomes the challenge caused by modality heterogeneity via learning modality-invariant and modality-specific representations, thus aiding the subsequent fusion for prediction tasks.
    \item We explicitly train the generated noise in a novel way to improve the quality of learned representations.
    To the best of our knowledge, we are the first work to model uninformativeness for a better shared-private disentanglement in MSA.
    \item MInD outperforms previous state-of-the-art methods on several standard multi-modal benchmarks only with a simple fusion strategy, which demonstrates the power of MInD in capturing diverse facets of multi-modal information.
\end{itemize}

\section{RELATED WORKS}

\subsection{Multi-modal Sentiment Analysis}

Learning effective joint representations is a critical challenge in MSA.
Many previous works have contributed to sophisticated fusion techniques. 
Zadeh et al.~\cite{zadeh2017tensor} proposed tensor-based fusion network which applies outer product to model the uni-modal, bimodal and tri-modal interactions.
Mai et al.~\cite{mai2020modality} introduced graph fusion network which regards each interaction as a vertex and the corresponding similarities as weights of edges.
Besides, the attention mechanisms~\cite{vaswani2017attention} are widely used to identify important information~\cite{shenoy2020multilogue,akhtar2019multi,lu2016hierarchical}. 
For instance, Tsai et al.~\cite{tsai2019multimodal} developed a novel transformer architecture that effectively integrates unaligned data from different modalities by directional pairwise cross-modal attention.
Shenoy et al.~\cite{shenoy2020multilogue} assigned weights to the importance differences between multiple modalities through the importance attention network.
Delbrouck et al.~\cite{delbrouck2020transformer} utilized a Transformer-based joint-encoding (TBJE) model, incorporating modular co-attention and a glimpse layer to effectively encode and analyze emotions and sentiments from one or more modalities.
However, as multi-modal inputs have various characteristics and information properties, this inherent heterogeneity of different modalities complicates the analysis of data, thus leading to a significant challenge on the mining and integration of information and the learning of multi-modal joint embedding.

\subsection{Disentanglement Learning}

Disentanglement learning~\cite{bousmalis2016domain,kim2018disentangling} is designed to unravel complex data structures, isolating key components to extract desirable information for more insightful and efficient data processing.
Therefore, this approach plays a pivotal role in aligning semantically related concepts across different modalities and effectively alleviates the problems caused by the modality gap. 
Furthermore, disentanglement learning significantly contributes to multi-modal fusion by offering a more structured and explicit representation~\cite{hazarika2020misa}. 
Such clarity and organization in the data representation are instrumental in enhancing the efficacy and precision of multi-modal integration processes.
For this reason, following Salzmann et al.~\cite{salzmann2010factorized}, many works have extended the shared-private learning strategies in various scenarios for excellent results, including retrieval~\cite{guo2019learning}, user representation in social network~\cite{tang2021learning}, and emotion recognition~\cite{yang2022disentangled}, etc.
In comparison, to the best of our knowledge, we provide the first attempt that explicitly train the generated noise in addition to the modality-invariant and modality-specific components for a better disentangled decomposition in MSA.

\section{METHOD}

\subsection{Model Overview}

\begin{figure}[t]
\centering
\includegraphics[width=0.95\textwidth]{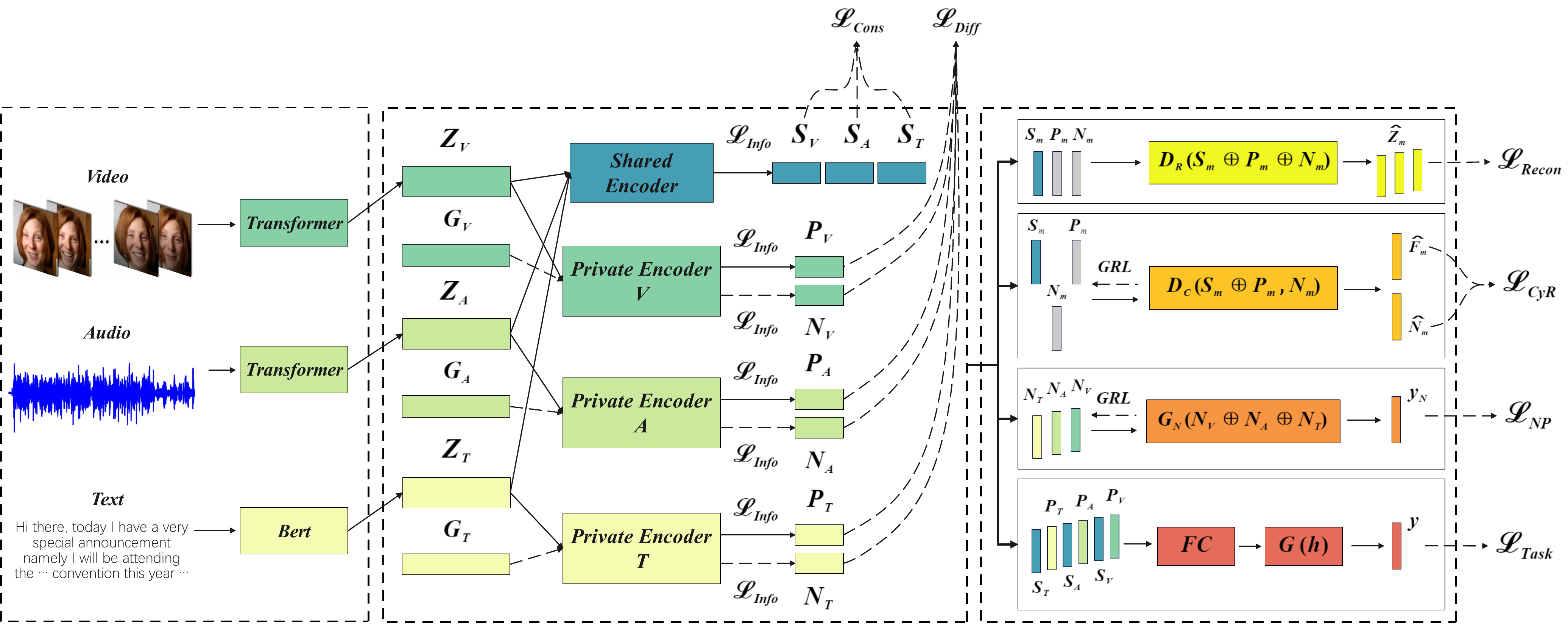}
\caption{Our proposed model. 
Each embedding $Z_{m}$ from the backbones is fed into a shared encoder and multiple private encoders to generate modality-invariant components $S_{m}$ and modality-specific components $P_{m}$, respectively. 
Meanwhile, by passing Gaussian noise $G_{m}$ to the private encoders, we align the uninformativeness within the feature subspace. 
This process is guided by mutual information maximization, the consistency loss and difference loss.
After that, we proposed the vanilla reconstruction module, the cyclic reconstruction module and the noise prediction module to further improve the representations.}
\label{fig:model}
\end{figure}

The overall framework of MInD is shown in Fig.\ref{fig:model}.  
We introduce our method within the context of a task scenario that incorporates three distinct modalities, namely visual, audio, and text.
Each individual data point consists of three sequences of low-level features originating from the visual, acoustic, and textual modalities. 
We denote them as 
${\bf X_{V}} \in \mathbf{R}^{L_{V} \times d_{V}}$,
${\bf X_{A}} \in \mathbf{R}^{L_{A} \times d_{A}}$,
${\bf X_{T}} \in \mathbf{R}^{L_{T} \times d_{T}}$,
respectively, where 
$L_{(\cdot)}$ 
is the sequence length and 
$d_{(\cdot)}$ 
is the embedding dimension.

In response to the challenges posed by modality heterogeneity, we aim to identify an approach that effectively mitigates the distributional discrepancy and enhances the information extraction, ensuring a comprehensive analysis of multi-modal inputs.
To this end, we decompose the inputs into two parts: the modality-invariant components and the modality-specific components.
The uninformativeness is modeled starting from generated Gaussian noise which is sent to the corresponding private encoder for feature subspace alignment, see details and explanation in the following subsections.
These representations are then facilitated through the implementation of information constraints, consistency constraints, and difference constraints. 
The integration of these constraints contributes to better utilization of information embedded within the high-level feature, enabling efficient exploration of both cross-modal commonality and distinctive features.
After that, we evaluate the completeness of decomposed information through a vanilla reconstruction module.
Moreover, we employ a cyclic reconstruction module to further reduce information redundancy, and a noise prediction module to minimize the task-related information within the trained noise.
The modality-invariant components and the modality-specific components are finally fused for prediction tasks.

\subsection{Feature Extraction}

Here we employ transformer-based models~\cite{vaswani2017attention} to extract high-level semantic features from individual modalities. 
Specifically, we use the Bert~\cite{devlin2018bert} model for text modality, and we employ a standard transformer model for the remaining two modalities, 
\begin{equation}
    Z_{m} = F^{m}(X_{m};{\theta}_{m}), \quad m \in \{V, A, T\}.
\end{equation}
The refined features of each modality are in a fixed dimension as 
$Z_{m} \in \mathbf{R}^{d_{k}}$.

\subsubsection{Modality-Invariant and -Specific Components.}
While temporal model-based feature extractors effectively capture the long-range contextual dependencies presented in multi-modal sequences, they fail to effectively handle feature redundancy due to the divergence of different modalities~\cite{zhang2022tailor}. 
Furthermore, the efficacy of the divide-and-conquer processing pattern is affected by the inherent heterogeneity among different modalities.

Inspired by these observations, we employ the shared and private encoders to learn the modality-invariant components and the modality-specific components, which are designed to capture commonality and specificity of individual modalities, respectively.
We denote the shared encoder as 
$E_{S}(\cdot;{\theta}_{S})$, 
and the private encoders as 
$E_{P_{m}}(\cdot;{\theta}_{P_{m}})$, where $m\in\{V, A, T\}$.
Then the representations are formulated as below:
\begin{equation}
S_{m} = E_{S}(Z_{m};{\theta}_{S}),
\end{equation}
\begin{equation}
P_{m} = E_{P_{m}}(Z_{m};{\theta}_{P_{m}}),
\end{equation}
\noindent with $S_{m}, P_{m}\in\mathbf{R}^{d_{k}}$.
The shared encoder $E_{S}(\cdot;{\theta}_{S})$ shares the parameters ${\theta}_{S}$ across all modalities, while the private encoders $E_{P_{m}}(\cdot;{\theta}_{P_{m}})$ assign separate parameters ${\theta}_{P_{m}}$ for each modality.
Both the shared encoder and private encoders are implemented as simple linear network with the activation function of GeLU~\cite{hendrycks2016gaussian}.

\subsubsection{Modeling Uninformativeness.}
While the integration of modality-invariant and modality-specific components facilitates a comprehensive representation of multi-modal inputs in former works~\cite{hazarika2020misa, yang2022disentangled}, we argue that this kind of approaches has not yet reached its full potential. 
The main problem lies in the persistence of meaningless information within the desired representations, for example, modality-specific unknown noise may be categorized as complementary components.
This may compromise information purity and limit the model's expressive capacity.
It is infeasible to directly isolate the noise from the modality-specific components as there lacks ground-truth nor any reference signal. 
Instead, drawing inspiration from adversarial training, we adopted an indirect approach for such separation. 
Our novel idea is that the private encoders should distinguish the informative input signals from the uninformativeness, mapping them into distinct areas. 
To this end, we first generate Gaussian noise vectors which are subsequently aligned into feature subspace using the same private encoders, namely:
\begin{equation}
    G_{m} \sim \mathcal{N}(0, 1), \quad
    N_{m} = E_{P_{m}}(G_{m};{\theta}_{P_{m}}),
\end{equation}
where $m\in\{V, A, T\}$.
We further require the outputs of a private encoder for the normal input $Z_{m}$ and the noise to be different through constraints, as described in next subsection.
This strengthens the robustness of the learned representations and aids in the extraction of more refined, purer information.

\subsection{Representation Objectives}

\subsubsection{Information Constraint.}
A conventional approach to discover useful representations involves maximizing the mutual information (MI) between the input and output of models. 
However, MI is notoriously difficult to compute, especially in continuous and high-dimensional contexts. 
Recent solution~\cite{hjelm2018learning} leverage mutual information constraint that estimate and maximize the MI between input data and learned high-level representations simultaneously.
Specifically, we employ the objective function based on the Jensen-Shannon divergence here, due to its proven stability and alignment with our primary aim of maximizing MI rather than obtaining an precise value.
The estimator is shown below:
\begin{align}
\mathcal{\hat{I}}_{\omega,\theta}^{(JSD)}(Z;E_{\theta}(Z)) 
= & E_{\mathcal{P}(Z,E_{\theta}(Z))}[-sp(-T_{\omega}(Z,E_{\theta}(Z)))] \nonumber \\
- & E_{\mathcal{P}(Z) \times \mathcal{P}(E_{\theta}(Z))}[sp(T_{\omega}(Z,E_{\theta}(Z)))],
\end{align}
where
$E_{\theta}(Z)$
is the encoder parameterized by $\theta$,
$\mathcal{P}(\cdot)$
is the empirical probability distribution,
$sp(\mathit{z}) = log(1+e^{\mathit{z}})$
is the softplus function, and
$T_{\omega} : \mathcal{X} \times \mathcal{Y} \rightarrow \mathbf{R}$
is a discriminator function modeled by a neural network with parameters $\omega$ called the statistics network.

Since the modality-invariant components are expected to capture cross-modal commonality, we calculate the MI between the outputs and the combination of inputs.
We also maximize the MI between the noise outputs and the generated Gaussian noise, encouraging $N_{m}$ to remain as less informative as the noise inputs after alignment through private encoders.
We denote the above procedure as following:
\begin{align}
\mathcal{L}_{Info} 
= & \sum_{m \in \{V,A,T\}} - \mathcal{\hat{I}}_{\omega_{S},\theta_{S}}^{(JSD)}(Z_{V} \oplus Z_{A} \oplus Z_{T};S_{m}) \nonumber \\
+ & \sum_{m \in \{V,A,T\}} - \mathcal{\hat{I}}_{\omega_{P_{m}},\theta_{P_{m}}}^{(JSD)}(Z_{m};P_{m}) \nonumber \\
+ & \sum_{m \in \{V,A,T\}} - \mathcal{\hat{I}}_{\omega_{P_{m}},\theta_{P_{m}}}^{(JSD)}(G_{m};N_{m}).
\end{align}

\subsubsection{Consistency Constraint.}
Inspired by~\cite{zbontar2021barlow}, we introduce the Barlow Twins loss (BT loss) to be the consistency constraint.
BT loss is originally designed for learning embedding which are invariant to distortions of the input sample, it forces two embedding vectors to be similar by making the cross-correlation matrix as close to the identity matrix as possible, which minimizes the redundancy between the components of these vectors.
Concretely, each representation pair $S^{A}, S^{B}$ is normalized to be mean-centered along the batch dimension as $S^{A,nor}, S^{B,nor}$, such that each unit has mean output 0 over the batch.
The normalized matrices can be then utilized to depict the cross-correlation matrix:
\begin{equation}
\mathcal{C}_{ij}^{A,B} = \frac{\sum_b s_{b,i}^{A,nor} s_{b,j}^{B,nor}}{\sqrt{\sum_b (s_{b,i}^{A,nor})^2}\sqrt{\sum_b (s_{b,j}^{B,nor})^2}},
\end{equation}
where $b$ indexes batch samples and $i, j$ index the vector dimension of the networks’ outputs.
The BT loss is expressed as:
\begin{equation}
\mathcal{L}_{BT}^{m_1,m_2} = \sum_{i} (1-\mathcal{C}_{ii}^{m_1,m_2})^2 
+ \lambda_{BT} \sum_{i}\sum_{j \ne i} (\mathcal{C}_{ij}^{m_1,m_2})^2.
\end{equation}

In the purpose of exploring shared information and commonality across modalities, we transfer the concept into our case by treating different modalities as different views.
Following the observations in~\cite{tsai2021note}, we set $\lambda_{BT}$ to be the dimension of the embedding, and calculate the BT loss between the modality-invariant components of each modalities pair:
\begin{equation}
\mathcal{L}_{Cons} = \sum_{(m_1,m_2)} \mathcal{L}_{BT}^{m_1,m_2}.
\end{equation}

\subsubsection{Difference Constraint.}
Since both the modality-invariant components and the modality-specific components are learned from the same high-level features $Z_{m}$, it may result in the redundancy of information. 
Moreover, as explained in former subsections, private encoders should effectively differentiate between informative and uninformative inputs.
For this sake, we employ the Hilbert-Schmidt Independence Criterion (HSIC)~\cite{song2007supervised} to measure independence.
Formally, the HSIC constraint between any two representations $R_{1}, R_{2}$ is defined as:
\begin{equation}
    HSIC(R_{1},R_{2}) = (n-1)^{-2} \textit{Tr} (UK_{1}UK_{2}),
\end{equation}
where $K_{1}$ and $K_{2}$ are the Gram matrices with
$k_{1,ij} = k_{1}(r_{1}^{i},r_{1}^{j})$
and
$k_{2,ij} = k_{2}(r_{2}^{i},r_{2}^{j})$.
$U = I - (1/n) e e^{T}$, 
where $I$ is an identity matrix and $e$ is an all-one column vector.
In our setting, we use the inner product kernel function for $K_{1}$ and $K_{2}$.
To augment the distinction among individual components, the overall difference constraint is expressed as:
\begin{equation}
    \mathcal{L}_{Diff} = \sum_{(R_1, R_2)} HSIC(R_1,R_2),
\end{equation}
where $(R_1, R_2)$ is the pair from $(S_m, P_m)$, $(S_m, N_m)$, $(P_{m_{1}}, P_{m_{2}})$, and $(P_m, N_m)$.

\subsubsection{Reconstruction Constraint.}
We adopt a vanilla reconstruction constraint, which aims to help the combination of representations capture more comprehensive information of their respective modality.
Note that we include $N_{m}$ during the reconstruction, as in our assumption that the modality-invariant components and the modality-specific components contain no meaningless information compared to the original signals.
By employing a decoder function
\begin{equation}
    \hat{Z}_{m} = D_{R}(S_m \oplus P_m \oplus N_m),
\end{equation}
the reconstruction constraint is then designed as the mean squared error between 
$Z_m$ and $\hat{Z}_{m}$:
\begin{equation}
    \mathcal{L}_{Recon} = \frac{1}{3} \sum_{m \in \{V,A,T\}} \frac{\| Z_m - \hat{Z}_{m} \|_{2}^{2}}{d_k},
\end{equation}
where $\| \cdot \|_{2}^{2}$ is the squared $L_2$-norm.

We further minimize the MI between the informative and uninformative vectors through cyclic-reconstruction and gradient-reversal layers~\cite{ganin2016domain}.
Let $F_{m}$ be the concatenation of $S_{m}$ and $P_{m}$.
$D_{C}(\cdot;\theta_{F_{m}})$, $D_{C}(\cdot;\theta_{N_{m}})$
be the decoders for reconstruction from $F_{m}$ to $N_{m}$ and from $N_{m}$ to $F_{m}$, respectively.
The objective is then formulated as below:
\begin{align}
\mathcal{L}_{CyR} 
= & \sum_{m\in\{V,A,T\}} \| F_{m} - D_{C}(GRL(N_{m});\theta_{N_{m}}) \|_{2}^{2} \nonumber \\
+ & \sum_{m\in\{V,A,T\}} \| N_{m} - D_{C}(GRL(F_{m});\theta_{F_{m}}) \|_{2}^{2},
\end{align}
where $GRL(\cdot)$ is a gradient reversal layer.

\subsection{Prediction}

Until now, the information disentanglement has been conducted in the unsupervised manner. 
We now complete our final objective function with the downstream task.
The learned modality-invariant and modality-specific components are first fused by a simple linear layer with dimension reduction, and subsequently trained via shallow MLPs: $G(\cdot;\theta_{G})$ with several hidden layers and GeLU activation to get the prediction denoted as
$\{\hat{y}_{i}\}$ or $\hat{Y}$:
\begin{equation}
    h = FC(S_V \oplus S_A \oplus S_T \oplus P_V \oplus P_A \oplus P_T),
\end{equation}
\begin{equation}
    \hat{Y} = G(h).
\end{equation}
Thanks to the explicit modeling of noise, the proposed disentangled decomposition allows for a fusion process that is simpler than alternative methods and results in improved performance. 

Specifically, to further reduce the task-related information inside the trained noise (thus the trained noise can be more meaningless in the sense of both informatics and task), we devise a noise-prediction loss with another shallow MLPs: $G_{N}(\cdot;\theta_{G_N})$, for the prediction $\{\hat{y}_{N,i}\}$ or $\hat{Y}_{N}$ from noise:
\begin{equation}
    \hat{Y}_{N} = G_{N}(GRL(N_V \oplus N_A \oplus N_T); \theta_{G_N}),
\end{equation}
\begin{equation}
\mathcal{L}_{NP} = -\frac{1}{n} \sum_{i=1}^{n} y_{i} \cdot log\hat{y}_{N,i} \quad
or \quad \frac{1}{n} \| Y - \hat{Y}_{N} \|_{2}^{2}.
\end{equation}
The final objective function is computed as:
\begin{align}
\mathcal{L}_{all} 
= & \mathcal{L}_{Task} 
+ \mathcal{L}_{NP} \nonumber
+ \alpha \mathcal{L}_{Info} \\
+ & \beta \mathcal{L}_{Cons}
+ \gamma \mathcal{L}_{Diff}
+ \lambda (\mathcal{L}_{Recon} + \mathcal{L}_{CyR}).
\end{align}
The seven loss terms are necessary and serve for different purposes, yet the final loss is controlled by only four hyper-parameters.
Here, $\alpha,\beta,\gamma,\lambda$ determine the contribution of each corresponding constraint to the overall loss.
And $\mathcal{L}_{Task}$ is the prediction loss, where we employ the standard cross-entropy loss for the classification task, and the mean error loss for the regression task.

\section{EXPERIMENTS}

\subsection{Datasets and evaluation criteria}

In this paper, we choose three multi-modal dataset for evaluation, namely CMU-MOSI and CMU-MOSEI for emotion recognition, and UR-FUNNY for humor detection. 

\subsubsection{CMU-MOSI.}
CMU-MOSI~\cite{zadeh2016multimodal} is a widely-utilized dataset for MSA. 
The dataset is collected from 2199 opinion video clips from YouTube, which is splited to 1284 samples for training set, 229 samples for validation set, and 686 samples for testing set, with sentiment score ranges from -3 to 3 for each sample.
Same as previous works, we adopt the 7-class accuracy ({Acc}-7), the binary accuracy ({Acc}-2), mean absolute error ({MAE}), the Pearson Correlation ({Corr}), and the {F}1 score for evaluation.

\subsubsection{CMU-MOSEI.}
CMU-MOSEI~\cite{zadeh2018multimodal} is a similar but larger dataset that contains 22,856 movie review video clips from YouTube, including 16,326 training samples, 1,871 validation samples, and 4,659 testing samples.
Each sample also has a sentiment scores ranging from -3 to 3. 
The same metrics are employed as in the above setting.

\subsubsection{URFUNNY.}
UR-FUNNY~\cite{hasan2019ur} dataset contains 16,514 samples of multi-modal punchlines labeled with a binary label for humor/non-humor instance from TED talks, which is partitioned into 10,598 samples in the training set, 2,626 in the validation set, and 3,290 in the testing set. 
We report the binary accuracy  ({Acc}-2) for this binary classification task.

\subsection{Implementation Details}

Following recent works, we utilize the pretrained {BERT-base-uncased} model to obtain a 768-dimension embedding for textual features.
Specifically, since the original transcripts are not available for our considered UR-FUNNY version, we follow the same procedure as~\cite{hazarika2020misa} to retrieve the raw texts from Glove~\cite{pennington2014glove}.
The acoustic features are extracted from COVAREP~\cite{degottex2014covarep}, where the dimensions are 74 for MOSI/MOSEI and 81 for UR-FUNNY.
Moreover, we use Facet\footnote{\url{https://imotions.com/platform/}} to extract facial expression features for both MOSI and MOSEI, and OpenFace~\cite{baltruvsaitis2016openface} for UR-FUNNY.
The final visual feature dimensions are 47 for MOSI, 35 for MOSEI, and 75 for UR-FUNNY.

Our model is built on the Pytorch 2.0.1 with one single Nvidia 3090 GPU. 
The number of transformer encoder layers for visual and audio are both 3. 
For the MOSI, MOSEI and UR-FUNNY benchmarks, the batch sizes and epochs are 32 and 100, respectively.

\subsection{Comparison With SOTA Models}

\begin{table}[t]
    \centering
    \begin{tabular}{l|ccccc|ccccc|c}
        \toprule
    \multirow{2}*{Models} & \multicolumn{5}{c|}{CMU-MOSI}  & \multicolumn{5}{c|}{CMU-MOSEI} & UR-FUNNY \\ 
    & \textit{Acc}7$\uparrow$ & \textit{Acc}2$\uparrow$ & \textit{F}1$\uparrow$ & \textit{MAE}$\downarrow$ & \textit{Corr}$\uparrow$ 
    & \textit{Acc}7$\uparrow$ & \textit{Acc}2$\uparrow$ & \textit{F}1$\uparrow$ & \textit{MAE}$\downarrow$ & \textit{Corr}$\uparrow$  
    & \textit{Acc}2$\uparrow$ \\
        \midrule
        TFN             & 34.9 & 80.8 & 80.7 & 0.901 & 0.698 & 50.2 & 82.5 & 82.1 & 0.593 & 0.700 & 68.57 \\
        LMF             & 33.2 & 82.5 & 82.4 & 0.917 & 0.695 & 48.0 & 82.0 & 82.1 & 0.623 & 0.677 & 67.53 \\
        MFM             & 35.4 & 81.7 & 81.6 & 0.877 & 0.706 & 51.3 & 84.4 & 84.3 & 0.568 & 0.717 & -     \\
        ICCN            & 39.0 & 83.0 & 83.0 & 0.862 & 0.714 & 51.6 & 84.2 & 84.2 & 0.565 & 0.713 & -     \\
        MulT            & 40.0 & 83.0 & 82.8 & 0.871 & 0.698 & 51.8 & 82.5 & 82.3 & 0.580 & 0.703 & -     \\
        Self-MM         & -    & 85.9 & 85.9 & 0.713 & 0.798 
                        & -    & 85.1 & 85.3 & 0.530 & 0.765 & -     \\
        HyCon           & 46.6 & 85.2 & 85.1 & 0.713 & 0.790
                        & 52.8 & 85.4 & 85.6 & 0.601 & 0.776 & -     \\
        BBFN            & 45.0 & 84.3 & 84.3 & 0.776 & 0.755 
                        & 54.8 & 86.2 & 86.1 & 0.529 & 0.767 & 71.68 \\
        CubeMLP         & 45.5 & 85.6 & 85.5 & 0.770 & 0.767 
                        & 54.9 & 85.1 & 84.5 & 0.529 & 0.760 & -     \\
        Liu et al.      & -    & 83.7 & 84.2 & 0.769 & 0.783 
                        & -    & 85.0 & 85.0 & 0.573 & 0.741 & -     \\
        AOBERT          & 40.2 & 85.6 & 86.4 & 0.856 & 0.700 
                        & 54.5 & 86.2 & 85.9 & 0.515 & 0.763 & 70.82 \\
        SURGM           & -    & 84.5 & 84.5 & 0.723 & 0.798 
                        & -    & 85.0 & 85.1 & 0.541 & 0.758 & -     \\
        ConFEDE         & 42.3 & 85.5 & 85.5 & 0.742 & 0.784 
                        & 54.9 & 85.8 & 85.8 & 0.522 & 0.780 & -     \\
        AcFormer        & 44.2 & 85.4 & 85.2 & 0.715 & 0.794
                        & 54.7 & 86.5 & 85.8 & 0.531 & 0.786 & -     \\
        TCHFN           & 44.8 & 86.1 & 86.3 & 0.748 & 0.780
                        & 53.2 & 86.3 & 86.5 & 0.538 & 0.770 & -     \\
        Self-HCL        & -    & 84.9 & 85.0 & 0.711 & 0.788
                        & -    & 85.9 & 85.9 & 0.531 & 0.775 & -     \\
        MISA            & 42.3 & 83.4 & 83.6 & 0.783 & 0.761 & 52.2 & 85.5 & 85.3 & 0.555 & 0.756 & 70.61 \\
        FDMER           & 44.1 & 84.6 & 84.7 & 0.724 & 0.788 
                        & 54.1 & 86.1 & 85.8 & 0.536 & 0.773 & 71.87 \\
        \midrule
    \textbf{MInD(ours)} & \textbf{46.6} & \textbf{86.0} & \textbf{86.0} & \textbf{0.711} & \textbf{0.791} 
                        & \textbf{53.9} & \textbf{86.6} & \textbf{86.7} & \textbf{0.529} & \textbf{0.772}  & \textbf{72.55} \\
        \bottomrule
    \end{tabular}
    \caption{Performance compared with the SOTA approaches in CMU-MOSI, CMU-MOSEI and UR-FUNNY, with MInD's results highlighted in bold.
    According to the comparison, while some baseline methods may stand out when evaluated by specific metric on one dataset, only MInD exhibits consistently competitive performance across all datasets under all metrics.}
    \label{tab:results}
\end{table}

\subsubsection{Baselines.}
We compare our model with many baselines, including pure learning based models such as
TFN~\cite{zadeh2017tensor},
LMF~\cite{liu2018efficient},
MFM~\cite{tsai2018learning}, and
MulT~\cite{tsai2019multimodal}.
Besides, we also compare our model with feature space manipulation approaches like
ICCN~\cite{sun2020learning},
MISA~\cite{hazarika2020misa},
Self-MM~\cite{yu2021learning},
HyCon~\cite{mai2022hybrid},
BBFN~\cite{han2021bi},
FDMER~\cite{yang2022disentangled} and
CubeMLP~\cite{sun2022cubemlp}.
Moreover, the more recent and competitive methods,
Liu et al.~\cite{liu2023improving},
AOBERT~\cite{kim2023aobert},
SURGM~\cite{hwang2023self},
ConFEDE~\cite{yang2023confede},
AcFormer~\cite{zong2023acformer},
TCHFN~\cite{hou2024tchfn} and
Self-HCL~\cite{fu2024self} are also taken into our consideration.
Results are directly taken from their corresponding paper.

\subsubsection{Multi-modal Emotion Recognition.}
As shown in Tab.\ref{tab:results}, MInD outperforms each baseline on most or even all evaluation metrics.
While some baseline methods may stand out when evaluated by specific metric on one dataset, only MInD exhibits consistently competitive performance across all datasets under all metrics.
Specifically, on the MOSI dataset, our approach shows the best results on {Acc}-7 and {MAE}, 
while on the MOSEI dataset, MInd surpasses all the SOTA {Acc}-2 and the {F}1 scores.
Although on MOSEI, {Acc}-7 of our approach is relatively lower than SOTA, it could be attributed to the fact that MInd only adopts simple concatenation and shallow linear network for fusion and prediction, which limits fine-grained sentiment calculation on larger dataset.
However, we still achieve overall satisfactory results without sophisticated fusion strategy, which reveals that our approach is able to capture sufficiently distinct information to form a comprehensive view of multi-modal inputs.
Notably, MInD significantly improves the performance on both datasets compared to MISA~\cite{hazarika2020misa} and FDMER~\cite{yang2022disentangled}, which are also disentanglement-based methods.
This is attributed to our introduction of trained noise that aids in the extraction of more refined, purer information through adversarial learning.

\subsubsection{Multi-modal Humor Detection.}
Further experiments are conducted on the UR-FUNNY dataset to verify the applicability of MInD.
Since humor detection is sensitive to heterogeneous representations of different modalities, the best
result achieved by MInD demonstrate the efficacy of our proposed multi-modal framework in learning distinct representations and capturing reliable information.

\subsection{Ablation Studies}

\begin{table}
    \centering
    \begin{tabular}{c|cc|cc|c}
        \toprule
    \multirow{2}*{Models} & \multicolumn{2}{c|}{CMU-MOSI}  & \multicolumn{2}{c|}{CMU-MOSEI} & UR-FUNNY \\ 
    & \textit{MAE}$\downarrow$ & \textit{Corr}$\uparrow$ 
    & \textit{MAE}$\downarrow$ & \textit{Corr}$\uparrow$  
    & \textit{Acc}2$\uparrow$ \\
        \midrule
    \textbf{MInD} & \textbf{0.711} & \textbf{0.791} 
                  & \textbf{0.529} & \textbf{0.772}
                  & \textbf{72.55} \\
        \midrule
    \multicolumn{6}{c}{Role of Modality} \\
        \midrule
    w/o visual & 0.857 & 0.771 & 0.541 & 0.770 & 71.12 \\
    w/o Audio  & 0.759 & 0.786 & 0.547 & 0.764 & 70.79 \\
    w/o Text   & 1.452 & 0.051 & 0.841 & 0.209 & 49.67 \\
        \midrule
    \multicolumn{6}{c}{Role of Disentanglement} \\
        \midrule
    w/o M-Invariant      & 0.793 & 0.778 & 0.546 & 0.767 & 70.30 \\
    w/o M-Specific       & 0.777 & 0.773 & 0.550 & 0.772 & 70.91 \\
    Non-Disentangled     & 0.925 & 0.753 & 0.576 & 0.772 & 68.97 \\
        \midrule
    \multicolumn{6}{c}{Role of Constraint} \\
        \midrule
    w/o $\mathcal{L}_{Info}$      & 0.755 & 0.778 & 0.542 & 0.761 & 71.22 \\
    w/o $\mathcal{L}_{Cons}$      & 0.789 & 0.777 & 0.551 & 0.760 & 72.28 \\
    w/o $\mathcal{L}_{Diff}$      & 0.768 & 0.769 & 0.556 & 0.762 & 71.70 \\
    w/o $\mathcal{L}_{Recon}$     & 0.727 & 0.784 & 0.558 & 0.758 & 72.01 \\
    w/o $\mathcal{L}_{CyR}$       & 0.787 & 0.773 & 0.539 & 0.763 & 72.28 \\
    w/o $\mathcal{L}_{NP}$        & 0.732 & 0.783 & 0.532 & 0.771 & 72.46 \\
    Only $\mathcal{L}_{Task}$     & 0.788 & 0.784 & 0.546 & 0.768 & 71.64 \\
        \bottomrule
    \end{tabular}
    \caption{Results of ablation studies.}
    \label{tab:ablations}
\end{table}

\subsubsection{Role of Modality.}
In Tab.\ref{tab:ablations}, we remove each modality separately to explore the performance of the bi-modal MInD, which performs consistently worse compared to the tri-modal MInD, suggesting that distinct modalities provides indispensable information.
Specifically, we observe a significant drop in performance when we remove the text modality, yet similar drops are not observed in the other two cases.
This shows the dominance of the text modality over the visual and audio modalities, probably due to the reason that the text modality contains manual transcriptions which could be inherently better, while on the contrary, the visual and audio modalities contain unfiltered raw signals with more noisy and redundant information.

\subsubsection{Role of Disentanglement.}
To empirically validate the effectiveness of the proposed disentanglement scheme, we carry out ablation studies on the modality-invariant components and the modality-specific components.
As shown in Tab.\ref{tab:ablations}, muting any one of the components leads to a degraded performance, indicating that each set of components capture different aspects of the information and is hence essential and meaningful.
In addition, we provide a non-disentangled version where the backbone features are directly utilized for fusion and prediction.
This situation shows even worse results on MOSI and UR-FUNNY, which further demonstrates the effectiveness of our approach.

\subsubsection{Role of Constraint.}
As shown in Tab.\ref{tab:ablations}, all the constraints show non-trivial contribution to the performance of MInD.
When there is no $\mathcal{L}_{Info}$, information extracted from the high-level features may be insufficient due to the adoption of simple shared and private encoders.
This in turn demonstrates that with the help of well designed constraints, neural network models can be simple yet effective.
When we remove $\mathcal{L}_{Cons}$ or $\mathcal{L}_{Diff}$, model fails to capture the shared information or specific information of distinct modalities.
In our model, $\mathcal{L}_{Recon}$ and $\mathcal{L}_{CyR}$ ensure the completeness and refinement of learned information, respectively. 
Removing them also brings worse performance.
The removal of $\mathcal{L}_{NP}$ leads to a slight degradation of performance on MOSEI and UR-FUNNY, and it is worth noting that the result on UR-FUNNY in this case still surpasses the baselines in Tab.\ref{tab:ablations}.
Finally, we present the results trained only with $\mathcal{L}_{Task}$.
The largest performance drop on most of the metrics demonstrates the necessity of all the constraints in our model.

\subsection{Visualization}

\begin{figure}[t]
    \centering
    \begin{subfigure}{0.9\textwidth}
        \centering 
        \includegraphics[width=0.3\textwidth]{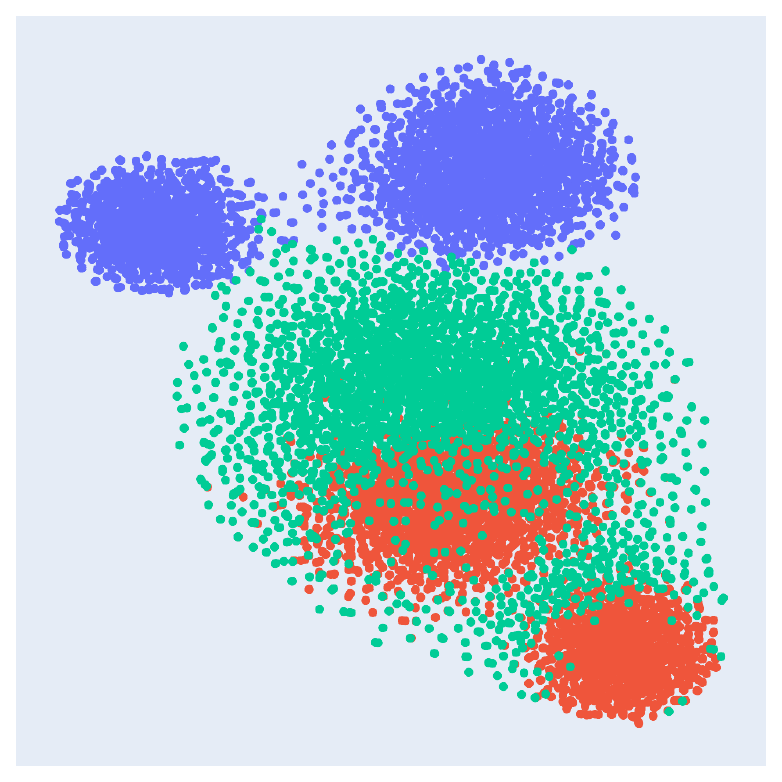}
        \includegraphics[width=0.3\textwidth]{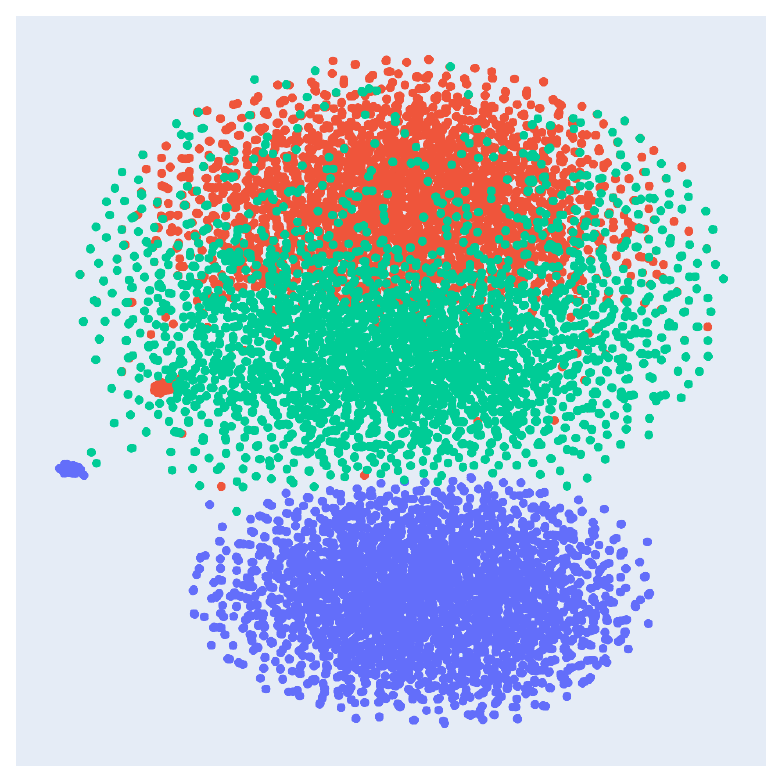}
        \includegraphics[width=0.3\textwidth]{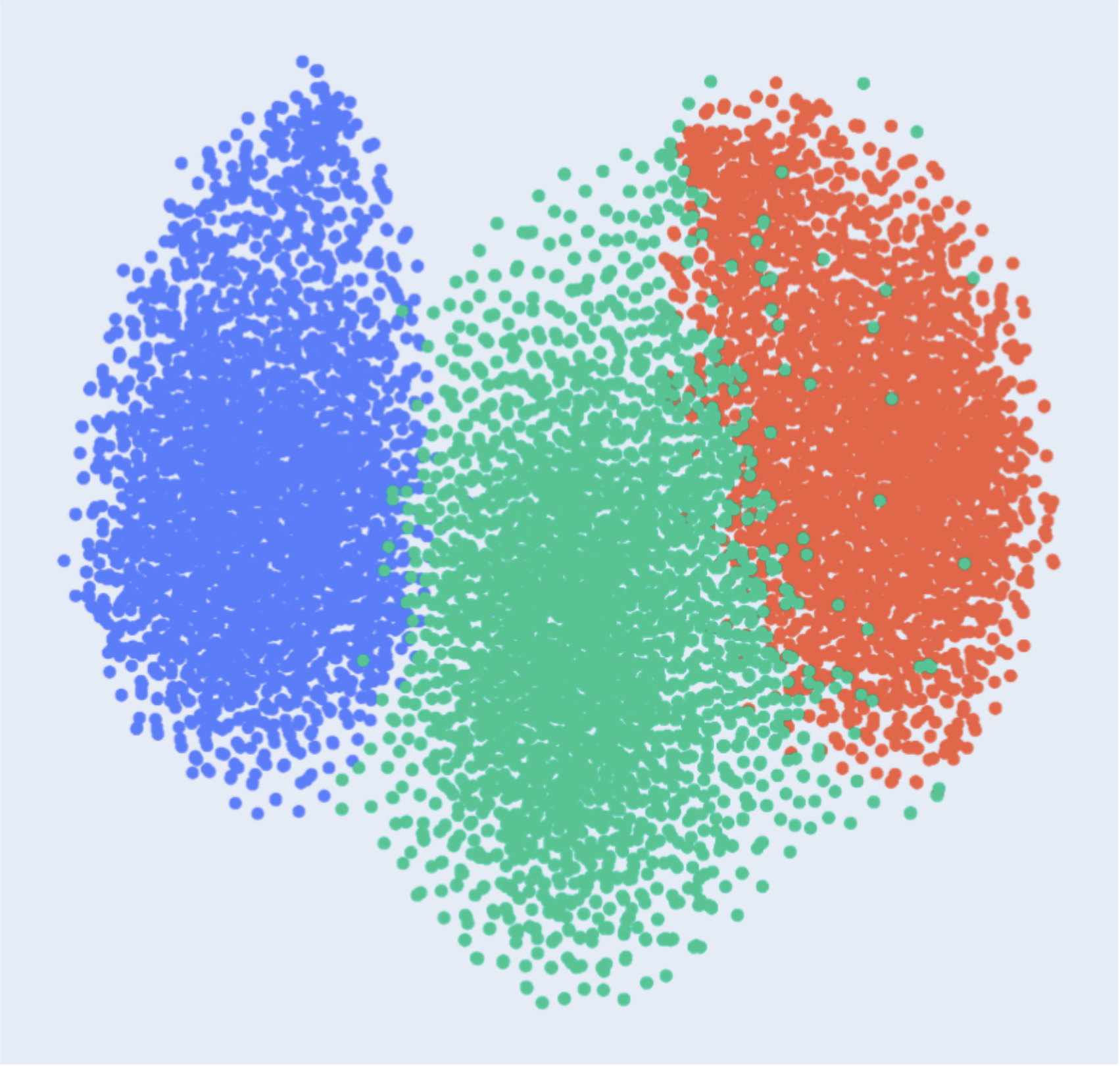}
        \caption{Visual, audio and text vectors before training, from left to right, respectively.}
        \label{fig:beforetrain}
    \end{subfigure}
    \begin{subfigure}{0.9\textwidth}
        \centering
        \includegraphics[width=0.3\textwidth]{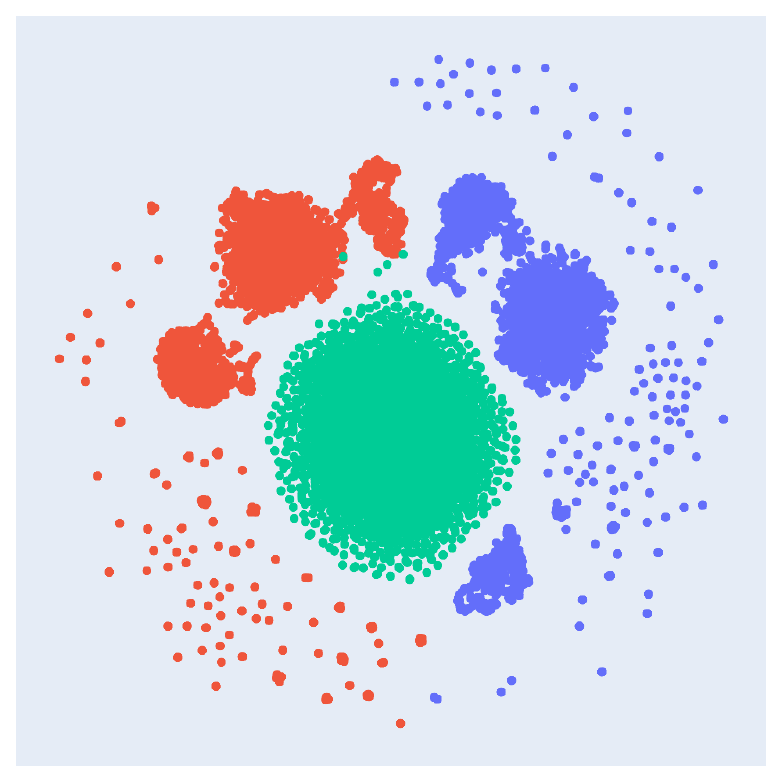}
        \includegraphics[width=0.3\textwidth]{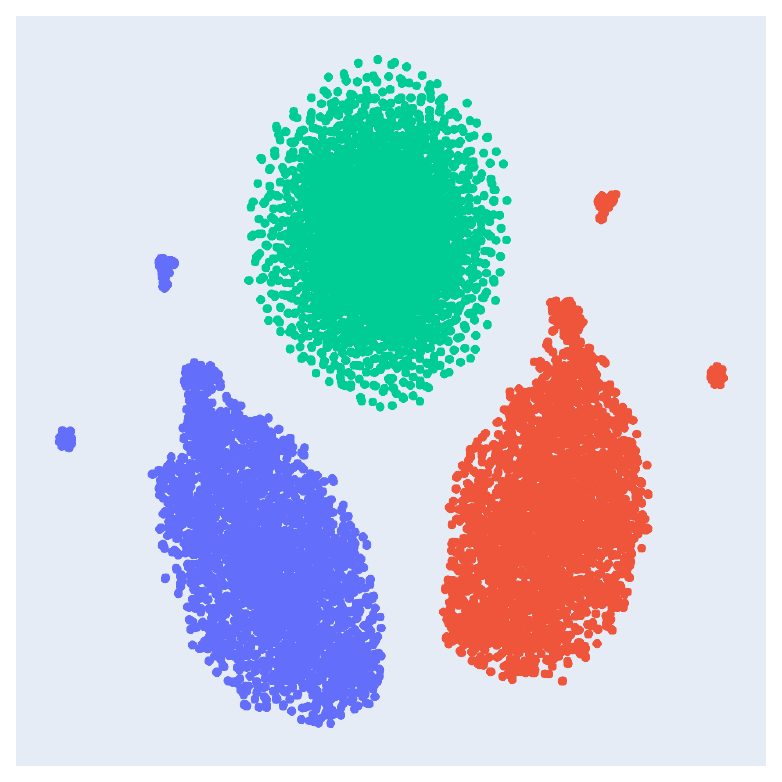}
        \includegraphics[width=0.3\textwidth]{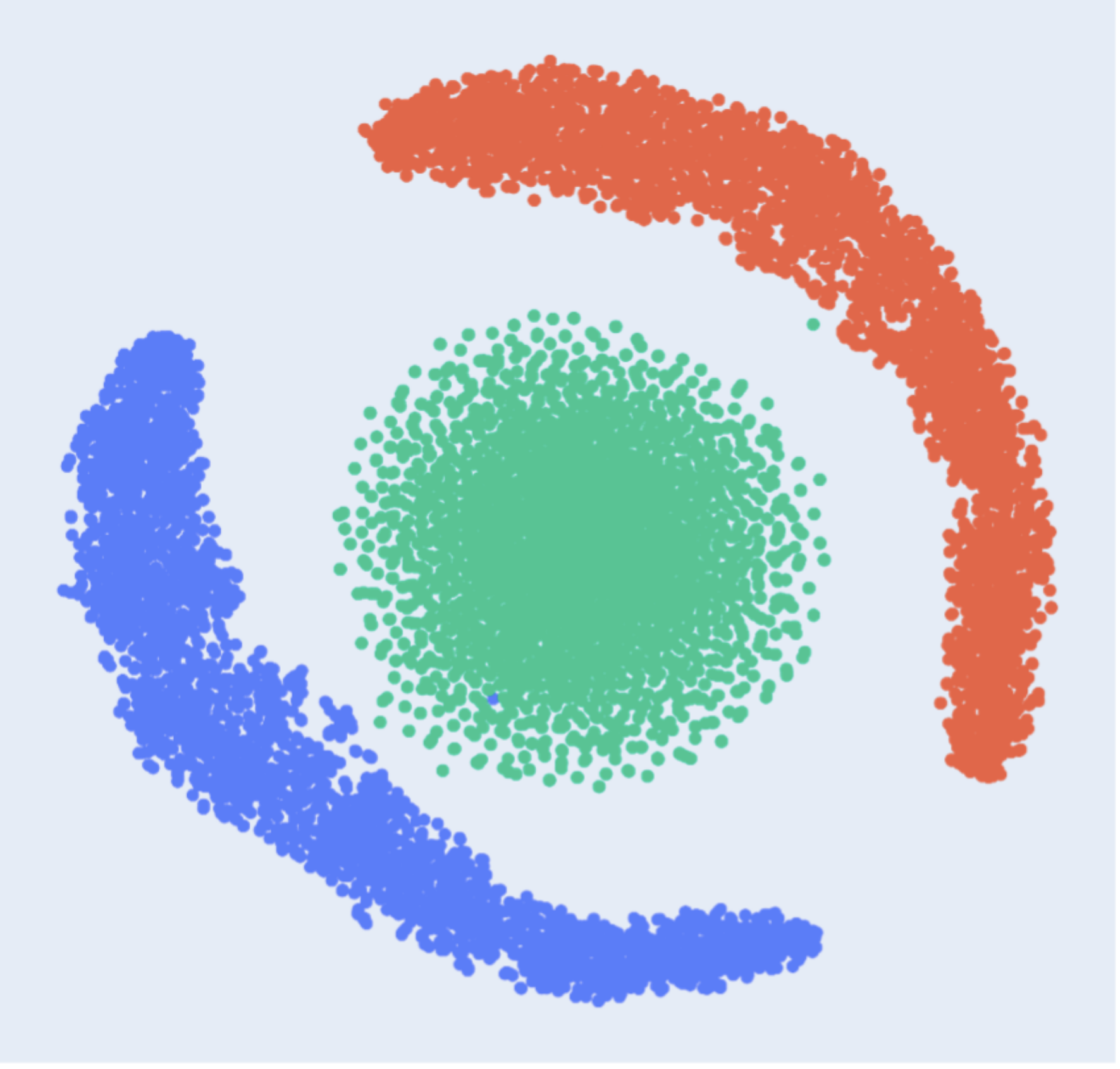}
        \caption{Visual, audio and text vectors after training, from left to right, respectively.}
        \label{fig:aftertrain}
    \end{subfigure}
    \caption{We present the visualization results of noise vectors, as well as the modality-invariant and modality-specific components from distinct modalities, taking the testing set of UR-FUNNY as example.
    Blue: Invariant; Red: Specific; Green: Noise.
    MInD is able to depict different aspects of information.}
    \label{fig:visualize}
\end{figure}

We visualize the noise vectors $N_{m}$, which are generated from Gaussian noise and subsequently aligned within the feature subspace by private encoders, along with the feature distributions of modality-invariant components $S_{m}$ and modality-specific components $P_{m}$ from individual modalities before and after training through T-SNE~\cite{van2008visualizing}, using data from the testing set of UR-FUNNY.

As shown below, before training, there is no clear boundary between the modality-specific components and the synthetic noise.
After training, the representations of the modality-specific components and the synthetic noise become more separated in the feature subspace, which helps a better exploitation of useful information.
This illustrates the potential of MInD in isolating the meaningless information, thus facilitating better representations.

\subsection{Experiment On Multi-modal Intent Recognition}

In order to investigate the generality of our proposed framework on other affective computing tasks, we carry out further experiment on MIntRec~\cite{zhang2022mintrec} dataset for multi-modal intent recognition.

MIntRec includes 2,224 high-quality samples, divided into 1,334 for training, 445 for validation, and 445 for testing. 
This dataset is designed to categorize intents into two levels: coarse-grained and fine-grained. 
The coarse-grained level features binary intent labels, differentiating between expressing emotions or attitudes and pursuing goals. 
The fine-grained level provides a more detailed classification, with 20 intent labels: 11 related to expressing emotions or attitudes and 9 focused on goal achievement.

We compare MInD's performance on the more difficult fine-grained level with the state-of-the-art multi-modal intent recognition methods, including MulT~\cite{tsai2019multimodal}, MISA~\cite{hazarika2020misa}, MAG-BERT~\cite{rahman2020integrating}, SPECTRA~\cite{yu2023speech} and CAGC~\cite{sun2024contextual}.
The results are directly taken from their corresponding paper.
As shown below, our model has achieved SOTA performance.

\begin{table}[h]
    \centering
    \begin{tabular}{ccccc}
        \toprule
        Methods   & Acc-20  & F1     & Pre.  & Rec.  \\
        \midrule
        MulT      & 72.52   & 69.25  & 70.25 & 69.24 \\
        MISA      & 72.29   & 69.32  & 70.85 & 69.24 \\
        MAG-BERT  & 72.65   & 68.64  & 69.08 & 69.28 \\
        SPECTRA   & 73.48   & -      & -     & -     \\
        CAGC      & 73.39   & 70.09  & 71.21 & 70.39 \\
        \textbf{MInD(Ours)} & \textbf{73.71}  & \textbf{70.12}  & \textbf{72.34} & \textbf{69.66} \\
        \bottomrule
    \end{tabular}
    \label{tab:mintrec}
    \caption{Performance on MIntRec Dataset.
    Our results are highlighted in bold.}
\end{table}

\section{CONCLUSION}

In this paper, we propose the Multi-modal Information Disentanglement (MInD) method to overcome the challenges caused by the inherent heterogeneity of distinct modalities through the decomposition of multi-modal inputs into modality-invariant and modality-specific components.
We obtain the refined representations via the well-designed constraints and improve the quality of disentanglement with the help of explicitly training the generated noise in an adversarial manner, which provides a new insight to pay attention to the meaningless information during the learning of different representation subspace.
Experimental results demonstrate the superiority of our method. 
In the future, we plan to broaden the application spectrum of our method, deploying it across a diverse array of multi-modal scenarios.

\bibliographystyle{unsrt}  
\bibliography{references}

\end{document}